\documentclass[aps,prb,twocolumn,superscriptaddress,letterpaper]{revtex4-2}

\usepackage{graphicx}
\usepackage{bm}
\usepackage{amssymb}
\usepackage{color}
\usepackage{amsmath}
\usepackage[colorlinks=true,linkcolor=blue,anchorcolor=black,citecolor=blue,filecolor=black,menucolor=black,runcolor=black,urlcolor=blue]{hyperref}

\begin{document}

\title{Pseudomagnetic suppression of non-Hermitian skin effect}

\author{Hau Tian Teo}
\affiliation{Division of Physics and Applied Physics, School of Physical and Mathematical Sciences, Nanyang Technological University, Singapore 637371, Singapore}

\author{Subhaskar Mandal}
\affiliation{Division of Physics and Applied Physics, School of Physical and Mathematical Sciences, Nanyang Technological University, Singapore 637371, Singapore}

\author{Yang Long}
\affiliation{Division of Physics and Applied Physics, School of Physical and Mathematical Sciences, Nanyang Technological University, Singapore 637371, Singapore}

\author{Haoran Xue}
\email{haoran001@e.ntu.edu.sg}
\affiliation{Division of Physics and Applied Physics, School of Physical and Mathematical Sciences, Nanyang Technological University, Singapore 637371, Singapore}

\author{Baile Zhang}
\email{blzhang@ntu.edu.sg}
\affiliation{Division of Physics and Applied Physics, School of Physical and Mathematical Sciences, Nanyang Technological University, Singapore 637371, Singapore}
\affiliation{Centre for Disruptive Photonic Technologies, Nanyang Technological University, Singapore 637371, Singapore}

\begin{abstract}
It has recently been shown that the non-Hermitian skin effect can be suppressed by magnetic fields. In this work, using a two-dimensional tight-binding lattice, we demonstrate that a pseudomagnetic field can also lead to the suppression of the non-Hermitian skin effect. With an increasing pseudomagnetic field, the skin modes are found to be pushed into the bulk, accompanied by the reduction of skin topological area and the restoration of Landau level energies. Our results provide a time-reversal invariant route to localization control and could be useful in various classical wave devices that are able to host the non-Hermitian skin effect but inert to magnetic fields.
\end{abstract}

\maketitle

\section{Introduction}

Due to the absence of Hermiticity, non-Hermitian systems can exhibit many unprecedented phenomena without Hermitian counterparts \cite{el2018non, ashida2020non, bergholtz2021exceptional, bender1998real, heiss2012physics, chong2010coherent, lin2011unidirectional, doppler2016dynamically, chen2017exceptional, yao2018edge}. The inherent point-gap topology in non-Hermitian systems has led to the emergence of non-Hermitian skin effect (NHSE), where an extensive number of eigenstates accumulate at the boundaries \cite{yao2018edge, alvarez2018, xiong2018does, gong2018topological, kunst2018biorthogonal}. Associated with the failure of conventional Bloch band theory and the breakdown of bulk-boundary correspondence in topological systems \cite{yao2018edge, kunst2018biorthogonal}, NHSE has been successfully observed in a few platforms \cite{brandenbourger2019non, helbig2020generalized, xiao2020non, liu2021non, zou2021observation, chen2021realization, zhang2021observation, zhang2021acoustic, gao2022non, gu2022transient}, leading to potential applications in wave manipulation, lasing and sensing \cite{weidemann2020topological, longhi2018non, zhu2022anomalous, teo2022, budich2020non, mcdonald2020exponentially, mandal2022topological}.
Contrary to boundary-localized skin modes, a magnetic field can induce Landau levels with eigenstates that are localized in the bulk \cite{landau2013quantum, klitzing1980new, thouless1982quantized}. Recent theoretical studies show that magnetic fields can lead to a suppression of the NHSE: with an increasing magnetic field, the skin modes are gradually pushed into the bulk \cite{lu2021magnetic, shao2022cyclotron}. Based on magnetic fields that induce bulk-localized Landau levels, the time-reversal breaking nature seems to be intrinsic to the suppression of boundary-localized NHSE.

By adopting the similar motivation of spatially varying gauge fields, a time-reversal invariant partner of magnetic field has also been widely studied, namely the pseudomagnetic field (PMF). Interestingly, using the PMFs which are artificially constructed from spatially inhomogeneous gauge fields, Landau levels and the associated bulk-localized modes can even be induced in magnetic-free systems without breaking time-reversal symmetry \cite{guinea2010energy}. This idea has been adopted in various classical wave systems to mimic magnetic-like effects and, in particular, to realize bulk-localized Landau modes in time-reversal invariant systems \cite{rechtsman2013strain, yang2017strain, abbaszadeh2017sonic, wen2019acoustic, jamadi2020direct, bellec2020observation, wang2020moire, guglielmon2021landau, yan2021pseudomagnetic, xue2020non}.

Therefore, the introduction of PMF gives rise to a potential alternative to suppress the NHSE without breaking time-reversal symmetry. In addition, while magnetic field takes effect in systems with charged particles like electrons, current platforms for realizing the NHSE are mostly magnetically inert \cite{brandenbourger2019non, helbig2020generalized, xiao2020non, liu2021non, zou2021observation, chen2021realization, zhang2021observation, zhang2021acoustic, gao2022non}. This separation in physical systems poses a challenge in utilizing the competition between the NHSE and magnetic field. By hosting similar localization mechanisms as magnetic fields \cite{lu2021magnetic, shao2022cyclotron}, a more feasible implementation of PMF suggests a novel type of suppression of the NHSE, thus offering a promising route for wave localization control in systems inert to magnetic fields.

\begin{figure}[b]
  \centering
  \includegraphics[width=0.8\columnwidth]{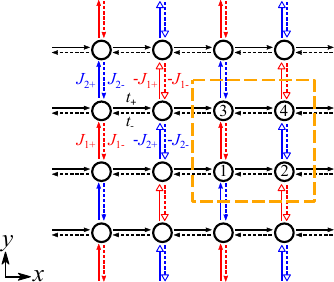}
  \caption{Illustration of the lattice model. The solid and dashed arrows denote hoppings along opposite directions, with the colors (i.e., black, red and blue) indicating different hopping strengths. Due to the presence of negative hoppings, there is a $\pi$ gauge flux per plaquette. The yellow dashed box denotes the unit cell containing four sites with their sublattice indices.}
  \label{fig1}
\end{figure}

In this work, we show that, instead of a real magnetic field, a PMF can also suppress the NHSE. We construct a two-dimensional lattice model with both NHSE and a PMF, induced by nonreciprocal hoppings and inhomogeneous hoppings, respectively. The NHSE can occur in either $x$ or $y$ direction, depending on where the nonreciprocal hoppings are implemented. We find that, for both cases, the NHSE will be suppressed when the PMF is introduced, as revealed by the calculations of skin mode profiles, skin topological areas and Landau level spectra. The suppression is prominent for energies within the first few Landau levels, where the effective theory of the PMF stays valid. Moreover, our model can be directly mapped to several realistic physical settings in photonic, acoustic and circuit systems, therefore paving a novel way to localization control for classical wave devices without breaking time-reversal symmetry.

\section{Generation of the PMF}

We consider a two-dimensional lattice model with $\pi$ flux in each plaquette, as illustrated in Fig.~\ref{fig1}. This lattice consists of nonreciprocal hoppings $t_\pm=t\pm\delta_x$ along $x$ direction, together with nonreciprocal and dimerized hoppings $J_{1\pm} = J_1\pm \delta_y$, $J_{2\pm} = J_2\pm \delta_y$ along $y$ direction. Half of the $y$-directional hoppings are set to be negative (i.e., $-J_{1\pm}$ and $-J_{2\pm}$), which introduce the $\pi$ flux. Note that all the hopping parameters are real. The Bloch Hamiltonian of this lattice is
\begin{equation}
	\mathcal{H}=
	\begin{bmatrix}
		0 & T_x+\Delta_x & V+\Delta_y & 0\\
		T_x+\Delta_x & 0 & 0 & -(V^\ast+\Delta_y)\\
		V^\ast+\Delta_y & 0 & 0 & T_x+\Delta_x\\
		0 & -(V+\Delta_y) & T_x+\Delta_x & 0
	\end{bmatrix} \label{eqHNH},
\end{equation}
where $T_x=2t\cos{(k_x/2)}$, $V=J_1 e^{ik_y/2}+J_2 e^{-ik_y/2}$, $\Delta_x=-2i\delta_x\sin{(k_x/2)}$ and $\Delta_y=-2i\delta_y\sin{(k_y/2)}$.
Without loss of generality, we set the lattice constant as $a=1$ thereafter and fix hopping amplitudes $t$ and $J_2$ to be $1$ and $0.5$, respectively. As we show below, the hopping dimerization and nonreciprocity can be used to generate a PMF and the NHSE, respectively. Therefore, we adopt this lattice to study the interplay between these two effects. 

\begin{figure}[b]
  \centering
  \includegraphics[width=\columnwidth]{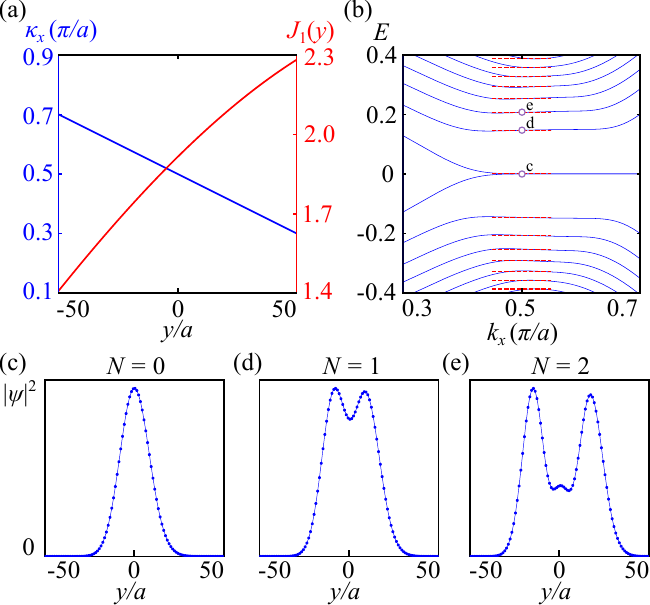}
  \caption{(a) Spatial distribution of hopping amplitude $J_1(y)$ in a strip of $N_y=100$ unit cells along $y$ direction with $\eta=0.2$, $t=1$ and $J_2=0.5$, parametrized by $\kappa_x$ in the construction of PMF in the Hermitian limit. (b) Partial view of the band diagram at $(\delta_x,\delta_y)=(0,0)$ and $\eta=0.2$, obtained from a strip with $N_y=100$ under $x$-PBC and $y$-OBC. Analytically derived Landau levels are labelled in red accordingly. Panels (c) to (e) show the intensity distributions of the first three Landau levels at $k_x=\pi/2$ labelled in panel (b).}
  \label{fig2}
\end{figure}

We first discuss the generation of the PMF in such a lattice. To this end, we let the non-Hermitian parameters $\delta_x$ and $\delta_y$ vanish. We note that the Hermitian limit of this model was studied in Refs.~\cite{shao2021gauge, xue2022projectively}, which focus on the consequences of projective symmetry algebra induced by the $\pi$ flux.

When $J_1$ and $J_2$ are equal, the model exhibits a four-fold band degeneracy at the corner of the Brillouin zone; when $J_1$ and $J_2$ differ, the four-fold degeneracy splits into two two-fold Dirac points at $(k_x,k_y)=(\tau\kappa_x,\pi)$, where $\kappa_x$ is related to the hoppings through $2t\cos{(\kappa_x/2)}=|J_1-J_2|$ and $\tau=\pm1$ is the valley index. To induce a uniform PMF, we introduce $y$-dependent hoppings $\pm J_1(y)$ of the following form:
\begin{equation}
	J_1(y) = J_2 + 2t\cos{\left[\frac12\left(\frac\pi2-\frac{2\pi y}{N_y}\eta \right)\right]}, \label{eqJ1}
\end{equation}
where $N_y$ is the number of unit cells along $y$ direction and $\eta$ is a dimensionless factor that controls the PMF strength for fixed $N_y$. Under such a spatially inhomogeneous hopping texture, the position of the Dirac points varies linearly in space from $(1/2+\eta)\pi$ to $(1/2-\eta)\pi$ as $y$ changes from $-N_y/2$ to $N_y/2$  [see Fig.~\ref{fig2}(a)], yielding the gauge field:
\begin{equation}
A=(A_x,A_y) =(-\tau\frac{2\pi}{N_y}\eta y, 0)
\end{equation}
The corresponding PMF is
\begin{equation}
	\mathcal{B} =\nabla\times A= \tau\frac{2\pi}{N_y}\eta
\end{equation}
It is noteworthy that the field strength $\mathcal{B}$ has opposite signs at opposite valleys. This indicates that the system is still time-reversal invariant, agreeing with the property of the PMF.

To see the effects of the PMF, we calculate the dispersion around $\tau=+1$ valley with $N_y=100$ and $\eta=0.2$. As shown in Fig.~\ref{fig2}(b), we can clearly see the formation of Landau levels, whose spacing matches the theoretical prediction elucidated in Appendix A:
\begin{equation}
	E_N =  \text{sgn}(N)\omega \sqrt{|N|}, N\in\mathbb{Z},\label{LL}
\end{equation}
where
\begin{align}
	\omega &= \sqrt{2 v_x v_y |\mathcal{B}|}
\end{align}
is the cyclotron frequency and $v_{x,y}$ is the group velocity. The eigenstates of Landau levels of order $N=0,1,2$ are plotted in Figs.~\ref{fig2}(c)-\ref{fig2}(e). As expected, they are all bulk-localized modes. By further increasing $\eta$, the eigenstates are further squeezed into the bulk due to a decreasing effective magnetic length, thus making the skin modes possible to be manipulated by tuning the strength of the PMF in our subsequent calculations.

\section{Generation of the NHSE}

After introducing the PMF, we set $\eta$ to zero to facilitate the generation of NHSE. In this limit, we tune the non-Hermitian components $\delta_x$ and $\delta_y$ to induce the NHSE along $x$ and $y$ directions, respectively. To investigate the NHSE, we conduct calculations under various boundary conditions. Henceforth, periodic boundary condition and open boundary condition along $x$($y$) direction are referred to as $x$($y$)-PBC and $x$($y$)-OBC, respectively.

\begin{figure}[b]
  \centering
  \includegraphics[width=\columnwidth]{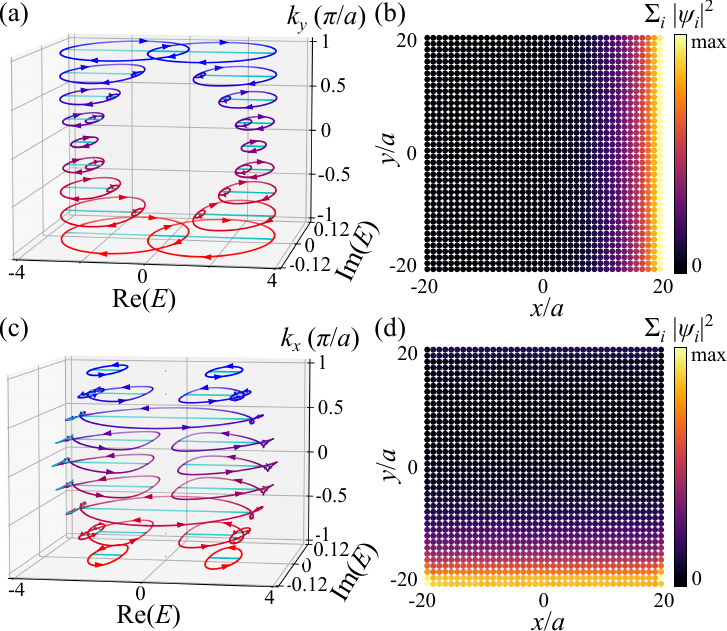}
  \caption{(a) Complex energy spectra at different $k_y$'s while considering $x$-PBC and $x$-OBC at $(\delta_x, \delta_y)=(0.05,0)$. By fixing $y$-PBC, those winding directions indicate the increasing direction of $k_x$ under $x$-PBC, whereas energy eigenvalues under $x$-OBC are plotted in cyan. (c) Complex energy spectra at different $k_x$'s while considering $y$-PBC and $y$-OBC at $(\delta_x, \delta_y)=(0,-0.05)$. By fixing $x$-PBC, those winding directions indicate the increasing direction of $k_y$ under $y$-PBC, whereas energy eigenvalues under $y$-OBC are plotted in cyan. Panels (b) and (d) depict the sum of intensity distribution of all eigenstates in a finite lattice with $40\times40$ unit cells, with non-Hermitian components in panels (a) and (c), respectively.}
  \label{fig3}
\end{figure}

We first illustrate the NHSE along $x$ direction by setting $(\delta_x, \delta_y)=(0.05,0)$. In this case, the eigenvalues under $x$-PBC and $y$-PBC form closed loops in the complex plane for each fixed $k_y$ [Fig.~\ref{fig3}(a)]. The spectral winding of the eigenvalues when $k_x$ increases from $-\pi$ to $\pi$ can be captured by the winding number (for a fixed $k_y$): 
\begin{equation}
	w(E_0) = \frac{1}{2\pi i} \int_{-\pi}^\pi dk_x\, \frac{d}{dk_x} \log{\det{(\mathcal{H}(k_x)-E_0)}}, \label{eqwe}
\end{equation}
where $E_0$ is a reference energy in the complex plane. This winding number is a topological invariant for point-gap topology and a nonzero $w$ indicates the emergence of NHSE under OBC \cite{okuma2020topological, zhang2020correspondence, borgnia2020non}. To see the NHSE, we plot the eigenvalues under $x$-OBC and $y$-PBC as cyan lines in Fig.~\ref{fig3}(a), which form open arcs in the interior of $x$-PBC and $y$-PBC spectra (the corresponding eigenmodes are skin modes). Therefore, for each fixed $k_y$, the corresponding 1D sub-system is a typical NHSE system. Subsequently, while considering a finite lattice (i.e., under $x$-OBC and $y$-OBC), we can find that the eigenstates are concentrated at the right boundary [Fig.~\ref{fig3}(b)].

The NHSE along $y$ direction can be similarly induced by letting $(\delta_x, \delta_y)=(0,-0.05)$. Under this condition, the eigenvalues under $x$-PBC and $y$-PBC form closed loops in the complex plane for each fixed $k_x$ [Fig.~\ref{fig3}(c)]. Consequently, the skin modes are now localized at the bottom boundary in a finite lattice, as shown in Fig.~\ref{fig3}(d).

\section{Competition between the PMF and the NHSE}

Comparing the PMF-induced Landau modes and the skin modes (see Figs.~\ref{fig2} and \ref{fig3}), it is evident that they have distinct localization areas: the Landau modes are localized in the bulk, while the skin modes are concentrated at the boundaries. In the present model, these two types of modes are induced and controlled by independent parameters, i.e., the PMF strength parameter $\eta$ and the nonreciprocal hopping parameters $(\delta_x, \delta_y)$. Next, we turn on these parameters simultaneously to study the competition between the two localization mechanisms and to reveal the suppression of the NHSE by the PMF. Note that the introduction of the PMF breaks the translational symmetry along $y$ direction, but still leaves $k_x$ a good quantum number. Therefore, we use different methods to investigate the cases when the NHSE is along $x$ and $y$ directions.

\subsection{Pseudomagnetic suppression of the NHSE along $x$ direction}

\begin{figure*}
  \centering
  \includegraphics[width=\linewidth]{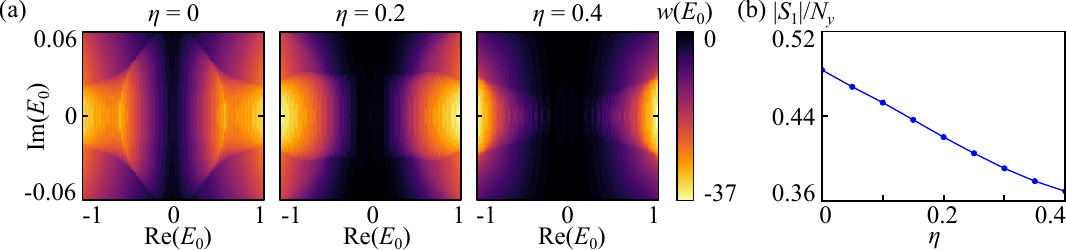}
  \caption{(a) Plots of winding number $w(E_0)$ against PMF indicator $\eta$ in the complex reference energy plane $[\text{Re}(E_0),\text{Im}(E_0)]$ at $(\delta_x, \delta_y)=(0.05,0)$. The color indicates the magnitude of winding number $w(E_0)$ obtained from a strip of $N_y=100$ unit cells along $y$ direction, but infinite along $x$ direction. (b) Skin topological area per unit cell $S_1/N_y$ as a function of PMF indicator $\eta$, computed in a similar supercell as panel (a).}
  \label{fig4}
\end{figure*}

When the NHSE is along $x$ direction, we use a bulk probe, i.e., the winding number defined in Eq.~\eqref{eqwe}, to detect the influence of the PMF on the NHSE. Using this method, we can avoid adopting $x$-OBC geometry where the computation is heavy and identification of modes from the bulk Dirac cones is relatively hard. In the calculation, a semi-infinite strip with $N_y=100$  along $y$ direction is considered, together with $x$-PBC and $y$-OBC imposed. In such a setup, the system can be regarded as a 1D lattice with a large number of sites in one unit cell. Thus, we can straightforwardly use the winding number to characterize the NHSE. The non-Hermitian parameters $(\delta_x, \delta_y)$ are fixed to be $(0.05,0)$, while the PMF strength parameter $\eta$ can be tuned. 

Figure~\ref{fig4}(a) shows the computed winding number in the complex plane for different PMF strength $\eta$. By gradually increasing $\eta$ from $0$ to $0.4$, regions of high winding number decrease and move away from zero energy. Winding number around zero energy even approaches zero at a stronger field, signifying the suppression of skin modes by the PMF. The effect of reduced winding number becomes weaker away from zero energy, consistent with the fact that the PMF is only valid in the low-energy regime. In general, the winding number in the entire complex plane decays when $\eta$ increases. To characterize the global behaviour of NHSE strength, skin topological area $S_1$ introduced in Ref.~\cite{lu2021magnetic} is employed here, which is the weighted sum of winding number in the entire complex energy plane. Figure~\ref{fig4}(b) highlights the relationship between $S_1$ (normalized by $N_y$) and $\eta$, showing that skin topological area decreases when PMF increases. This trend also demonstrates the pseudomagnetic suppression of the NHSE along $x$ direction.

\subsection{Pseudomagnetic suppression of the NHSE along $y$ direction}

\begin{figure}
  \centering
  \includegraphics[width=\columnwidth]{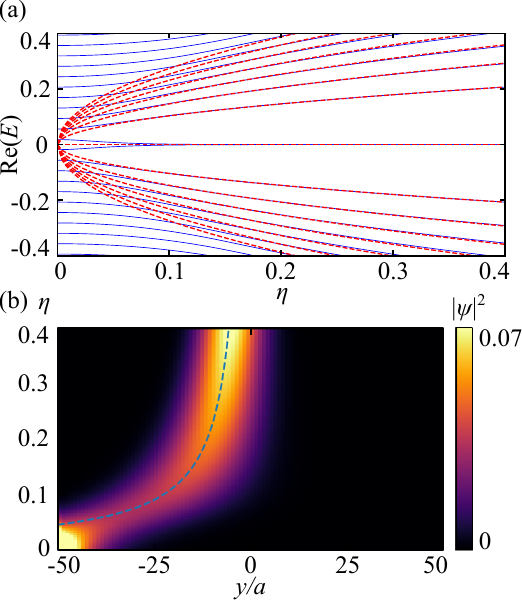}
  \caption{(a) Real part of the energy spectrum as a function of $\eta$ at the $k_x = \pi/2$ valley and fixed non-Hermitian parameters $(\delta_x, \delta_y)=(0,-0.05)$. Landau levels of order $N=-7$ to $7$ in the Hermitian limit are included as red dashed lines. (b) Spatial profile evolution of a zero-energy eigenstate in (a), showing a transition from a skin mode to a bulk-localized zeroth Landau mode as theoretically predicted by its center $y_0$ (blue dashed line).}
  \label{fig5}
\end{figure}

Now we turn to the other case where the NHSE is along $y$ direction. In this case, the winding number and skin topological area employed above will not be useful since they can only capture $x$-directional skin modes. However, we can directly access the skin modes by taking $x$-PBC and $y$-OBC while fixing $k_x$ at the valleys. Specifically, we set $k_x=\pi/2$ (i.e., the $\tau=+1$ valley) and $(\delta_x,\delta_y)=(0,-0.05)$ and adopt a semi-infinite strip ($N_y=100$) with $x$-PBC and $y$-OBC. In this setup, we are able to investigate in detail how the PMF will affect the NHSE.

The energy spectrum of the semi-infinite strip is first investigated near zero energy where the PMF description is expected to hold. In fact, the eigenenergies retrieve their Landau quantization values while PMF is introduced. This is numerically demonstrated in Fig.~\ref{fig5}(a), where blue curves show the eigenenergies (real part) computed from the lattice Hamiltonian at $k_x=\pi/2$, and red curves denote Landau levels predicted from Eq.~\eqref{LL}. As can be seen, the eigenenergies of the lowest few modes gradually approach Landau level energies as $\eta$ increases. As one of the signatures of the PMF, the quantization behaviour gives a hint of suppressing the NHSE using bulk-localized Landau modes.

By gradually tuning the PMF, the suppression is indeed observed in the eigenstates near zero energy. As shown in Fig.~\ref{fig5}(b), an eigenstate pinned to zero energy is chosen to visualize its spatial variation when $\eta$ increases. To be precise, this eigenstate hosts an eigenvalue characterized by the zeroth-order Landau level computed in the Hermitian limit. In the $\eta=0$ limit, the eigenstate displays an exponential decay from the boundary into the bulk, which is exactly a skin mode profile expected from an NHSE system. With an increasing $\eta$, this skin mode moves progressively into the bulk and forms a growing hump in the bulk. This mode profile evolution clearly demonstrates how the PMF can suppress the NHSE.

In addition, we can observe that the center of the mode moves continuously as $\eta$ varies, which could be a useful property for localization control. At a high PMF strength (e.g., $\eta=0.4$), the mode becomes very similar to a Landau mode with a Gaussian profile. As elucidated in Appendix A, its center is not exactly at the center of the lattice, but with a shifted value $y_0$. Under the presence of a nonzero $\delta_y$, $y_0$ can be described by non-Hermitian strength $\delta_y$ and PMF indicator $\eta$ as follows:
\begin{equation}
	y_0=\frac{N_y}{\pi v_x}\frac{\delta_y}{\eta} \label{eqy0}
\end{equation}
It is labelled as the blue dashed line in Fig.~\ref{fig5}(b), thus highlighting the competition between NHSE and PMF based on their indicators in Eq.~\eqref{eqy0}. Without loss of generality, the eigenstates characterized by higher-order Landau levels near zero energy [eigenvalues shown in Fig.~\ref{fig5}(a)] can also be conceptually described by the shifted center $y_0$, in other words, an additional tuning to the Landau mode solutions in the Hermitian limit. Consequently, these higher-order eigenstates can also be suppressed by the PMF scheme, showing a large tunability from skin modes to bulk modes [Fig.~\ref{fig6}].

\section{Conclusion and outlook}

To conclude, we have shown that the PMF can induce a suppression of the NHSE. In particular, a boundary-localized skin mode can be transferred into a bulk-localized mode by the PMF. We note that, while in the main text, we only show edge skin modes can be pushed into the bulk by the PMF. Corner skin modes actually behave similarly, as demonstrated in Appendix B. Compared to other methods to tune the NHSE, like using the real magnetic field \cite{li2022enhancement} and electric field \cite{peng2022manipulating}, the PMF approach would be easier to be engineered in various artificial systems. In Appendix C, we provide an electric circuit design of the tight-binding model, showing its feasibility in physical systems. Moreover, the PMF can reach higher values than the real magnetic field \cite{levy2010strain}. However, one needs to keep in mind that the PMF only holds for certain low-energy modes. Thus, not all skin modes can be well manipulated by the PMF. In future works, it would be desired to further apply the tunable skin modes found here to various classical wave devices, especially in optical systems where active control of localization and nonlinear effects can be studied.

\begin{acknowledgments}
This work is supported by Singapore Ministry of Education Academic Research Fund Tier 3 under Grant No. MOE2016-T3-1-006.
\end{acknowledgments}

\section*{Appendix A: PMF-induced Landau levels vs non-Hermiticity}

Based on the non-Hermitian Hamiltonian in Eq.~\eqref{eqHNH}, we will show that the PMF scheme in Eq.~\eqref{eqJ1} is able to generate Landau levels, at the same time suppressing the non-Hermitian skin effect. The Hermitian limit illustrated in Fig.~\ref{fig2} is naturally included in the subsequent derivations, by setting the non-Hermitian terms to zero.

We start from the most generic form of non-Hermitian Hamiltonian, followed by the eigenvalue equation $\mathcal{H}|\psi\rangle = E|\psi\rangle$. The Hamiltonian can be expressed in a block-diagonal form via a unitary transformation:
\begin{align}
	\mathcal{H}'=U^\dagger \mathcal{H} U =
	\begin{bmatrix}
		\mathcal{H}_0 & 0\\
		0 & \mathcal{H}_1
	\end{bmatrix}, U = \frac{1}{\sqrt2}
	\begin{bmatrix}
		1 & 0 & 0 & 1\\
		0 & 1 & 1 & 0 \\
		0 & i & -i & 0 \\
		i & 0 & 0 & -i
	\end{bmatrix}
\end{align}
with the following blocks $(W\equiv-iV)$:
\begin{eqnarray}
	\mathcal{H}_0 =\begin{bmatrix}
		0 & T_x+\Delta_x-W+i\Delta_y\\
		T_x+\Delta_x-W^\ast-i\Delta_y & 0
	\end{bmatrix} \nonumber \\
	\mathcal{H}_1 = 
	\begin{bmatrix}
		0 & T_x+\Delta_x+W^\ast+i\Delta_y \\
		T_x+\Delta_x+W-i\Delta_y & 0
	\end{bmatrix} \label{eqH0H1}
\end{eqnarray}
Upon this transformation, we observe the relation $\mathcal{H}' |\Phi\rangle = E |\Phi\rangle$, where $|\Phi\rangle \equiv U^\dagger|\psi\rangle=[|\Phi_0\rangle, |\Phi_1\rangle]^T$ is expressed in the two-component vectors $|\Phi_0\rangle$ and $|\Phi_1\rangle$. To elucidate the suppression of $y$-directional NHSE, we set $\delta_x=0$ throughout this section. Now, the Hamiltonian $\mathcal{H}'$ is expanded around $(k_x,k_y)=(\tau\pi/2+q_x,\pi+q_y)$ valleys $(\tau=\pm1)$ up to first order in $q_x$ and $q_y$:
\begin{align}
	T_x &= t\sqrt2(1-\tau \frac{q_x}{2}) + \mathcal{O}(q_x^2) \nonumber\\
	W &= -iV = (J_1-J_2) + i\frac{q_y}{2}(J_1+J_2) + \mathcal{O}(q_y^2)\\
	\Delta_y &= -2i\delta_y+\mathcal{O}(q_y^2) \nonumber
\end{align}

By observing that $\mathcal{H}_0 |\Phi_0\rangle = E |\Phi_0\rangle$, we arrive at the Dirac Hamiltonian with an additional imaginary gauge:
\begin{equation}
	\left[-\tau v_x \sigma_x q_x + (v_y q_y+2i\delta_y)\sigma_y\right]|\Phi_0\rangle = E|\Phi_0\rangle + \mathcal{O}(q^2)
\end{equation}
with group velocity $(v_x, v_y) = (t/\sqrt2, J_2+t/\sqrt2)$. The relations $J_1-J_2 = 2t\cos{(\kappa_x/2)}$ at $\kappa_x=\tau\pi/2$ valleys are utilized here, which correspond to the $\eta=0$ limit in Eq.~\eqref{eqJ1}. The $y$-dependent hoppings in Eq.~\eqref{eqJ1} are now introduced to the Dirac Hamiltonian up to first order in $y$, leading to the following Hamiltonian:
\begin{equation}
	\left[-\tau v_x\sigma_x (-i\partial_x -  A_x) - i v_y\sigma_y \partial_y + 2i\delta_y\right] \Phi_0(r) \simeq E \Phi_0(r) \label{eqHy}
\end{equation}
where gauge field $A_x$ is generated with respect to the reference Dirac point $\kappa_0^\tau\equiv \kappa_x^\tau(J_1 = J_2+t\sqrt2)=\tau\pi/2$:
\begin{equation}
	A_x(y) = \kappa_x^\tau [J_1(y)] - \kappa_0^\tau = -\tau\frac{2\pi}{N_y}\eta y
\end{equation}

The valley-dependent pseudomagnetic field $\mathcal{B}=\partial_x A_y - \partial_y A_x$ is thus obtained, with fundamental constants $a$, $\hbar$ and $e$ (electron charge) restored:
\begin{equation}
	\mathcal{B} = \tau\frac{2\pi}{N_y}\eta \stackrel{\text{restore }a,\hbar,e=1}{=} \tau\frac{2\pi\hbar}{N_y e a^2}\eta
\end{equation}

Since $J_1(y)$ is homogeneous along $x$ direction, $\Phi(r)$ can be modulated by a plane wave with $K_x= k_x - \kappa_0^\tau$, where $\Phi(r) = e^{iK_x x}\,\Phi(y)$:
\begin{equation}
	\left[-\tau v_x \sigma_x (K_x - A_x) + v_y (-i\partial_y+iq_y')\sigma_y\right]|\Phi_0\rangle \simeq E|\Phi_0\rangle
\end{equation}
where $q_y'=2\delta_y/v_y$. To remove the imaginary gauge, we redefine the eigenstate $|\Phi_0'\rangle=e^{-q_y' y}|\Phi_0\rangle$, arriving at:
\begin{align}
	\left[-\tau v_x \sigma_x (K_x-A_x) + v_y\sigma_y q_y\right]|\Phi_0'\rangle &\simeq E |\Phi_0'\rangle \label{eqHyNH}
\end{align}
This looks exactly like the Hermitian case, except with an additional similarity transformation on eigenstate. Therefore, Eq.~\eqref{eqHyNH} can be expressed via these coupled equations, where $\Phi_0'(r) = [\Phi_A'(r), \Phi_B'(r)]^T$:
\begin{align}
	[-\tau v_x (K_x-A_x) - v_y \partial_y] \Phi_B'(y) \simeq E \Phi_A'(y) \label{eqBANH}\\
	[-\tau v_x (K_x-A_x) + v_y \partial_y] \Phi_A'(y) \simeq E \Phi_B'(y) \label{eqABNH}
\end{align}
By combining Eq.~\eqref{eqBANH}-\eqref{eqABNH} with $\tau=+1$, we arrive at the eigenproblem in the form of quantum harmonic oscillator even in the non-Hermitian setup:
\begin{align}
	\omega^2 (a^\dagger a)\Phi_B' = E^2 \Phi_B', \label{eqQHONH}
\end{align}
where $\omega = \sqrt{t\sqrt2(2J_2+t\sqrt2)\frac{\eta\pi}{N_y}} = \sqrt{2v_x v_y |\mathcal{B}|}$, followed by annihilation operator $a = [v_x (K_x-A_x) + v_y \partial_y]/\omega$ and $[a,a^\dagger]=1$. Consequently, the eigenvalues $E^2$ employ the form of quantum harmonic oscillator, leading to the Landau level energies labelled by order $N$:
\begin{equation}
	E_N = \text{sgn}(N) \omega \sqrt{|N|}, N\in \mathbb{Z}
\end{equation}

As a result, the energy plateaus still exist in non-Hermitian case. We denote $A_x = -\mathcal{B}y$ as derived. For illustration in Fig.~\ref{fig5}, we set $K_x=0$ here. By expanding Eq.~\eqref{eqQHONH}, we arrive at:
\begin{align}
	\partial_y^2 \Phi_B' + \left[\epsilon^2 - \left(\frac{v_x}{v_y} \mathcal{B}\right)^2 y^2\right]\Phi_B' &= 0 
\end{align}
where $\epsilon^2 = (|N|+1/2)(\omega/v_y)^2$. This resembles the quantum harmonic oscillator problem, but with modified effective field $B_c$ that characterizes magnetic length $l_B$:
\begin{equation}
	B_c = \frac{v_x}{v_y}\mathcal{B},\,\,\, l_B^2 = B_c^{-1} = \frac{v_y}{v_x} \mathcal{B}^{-1}
\end{equation}
It is noteworthy that $B_c$ here is not exactly equal to $\mathcal{B}$ due to the intrinsic anisotropy of the tight-binding model without any strain. By observing Eq.~\eqref{eqQHONH}, we can deduce that at the zeroth Landau level $(E=0)$, the eigenstate employs the form: $\Phi_A'=0, \Phi_B'\propto e^{-y^2/2l_B^2}$. By recovering the eigenstate before imaginary gauge transformation $|\Phi_0\rangle  = [\Phi_A, \Phi_B]^T$, we obtain the final eigenstate:
\begin{align}
	\Phi_A=0, \Phi_B\propto e^{-y^2/2l_B^2}e^{q_y' y}\propto e^{-(y-y_0)^2/2l_B^2} \label{eqsol}
\end{align}
where shifted center $\displaystyle y_0=l_B^2 q_y'=\frac{N_y}{\pi v_x}\frac{\delta_y}{\eta}$. Since $y=y_0$ corresponds to the position with largest intensity, we can deduce that the $\delta_y/\eta$ term in $y_0$ describes the competition between non-Hermiticity $\delta_y$ and PMF $\eta$.

Due to the intrinsic sublattice symmetry in this model ($S\mathcal{H}S^{-1} = -\mathcal{H}$, $S = \sigma_z \otimes \sigma_z$), features of sublattice polarization are also captured here. Note that:
\begin{equation}
	\begin{bmatrix}
		\Phi_A \\
		\Phi_B \\
		\Phi_C \\
		\Phi_D
	\end{bmatrix} \equiv 
	\begin{bmatrix}
		|\Phi_0\rangle \\
		|\Phi_1\rangle
	\end{bmatrix} = U^\dagger|\psi\rangle = \frac{1}{\sqrt2}\begin{bmatrix}
		\psi_1 - i\psi_4\\ \psi_2 - i\psi_3 \\ \psi_2 + i\psi_3 \\ \psi_1 + i\psi_4
	\end{bmatrix},
\end{equation}
We can observe that $\Phi_A$ and $\Phi_B$ capture the features of $(\psi_1, \psi_4)$ and $(\psi_2,\psi_3)$ respectively. As a result, solution in Eq.~\eqref{eqsol} illustrates the complete distribution of the eigenstate in sublattices 2 and 3, in the form of Landau mode. This corresponds to the complete dominance of Landau modes at large $\eta$ in Fig.~\ref{fig5}(b). Regarding the transition from skin modes to Landau modes, it is not described in this model as the continuum model fails at the boundary. In other words, when $y_0$ is close to boundary $y=N_y/2$ (large $\delta_y$ or small $\eta$), the failure of the model leads to the emergence of skin modes.

\begin{figure}[t]
  \centering
  \includegraphics[width=\columnwidth]{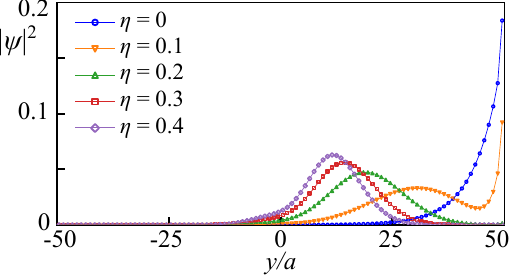}
  \caption{Spatial variation of an eigenstate characterized by first-order Landau level by increasing PMF $\eta$ at the $k_x=\pi/2$ valley and fixed non-Hermitian components $(\delta_x, \delta_y)=(0,0.05)$.}
  \label{fig6}
\end{figure}

In general, the arguments can be applied to the modes further from the valleys (i.e., $K_x \neq$ 0). It corresponds to the additional shift to $y_0$, which can be seen from Eq.~\eqref{eqHyNH}. Apart from zero-energy modes, the suppression is also prominent for the first few Landau modes (small $|N|$), falling in the region where PMF holds. The solutions are related to Hermite polynomials with an additional tuning by the exponential term $e^{q_y' y}$. The evolution of the eigenstate characterized by first Landau level is illustrated in Fig.~\ref{fig6} for a better visualization.

\section*{Appendix B: Manipulation of corner skin modes in a finite lattice}

\begin{figure}[b]
  \centering
  \includegraphics[width=\columnwidth]{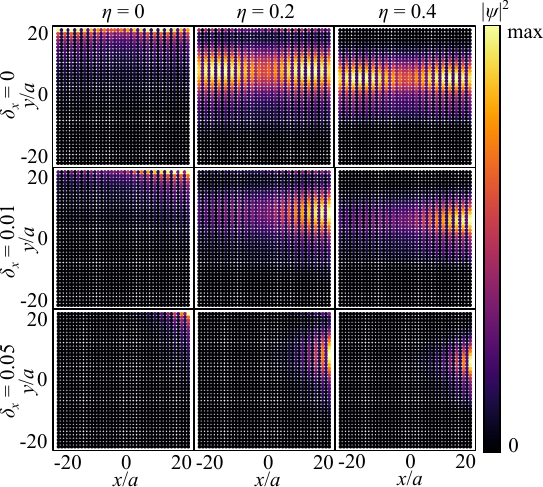}
  \caption{Spatial variation of an eigenstate near zero energy level in lattices with $40\times40$ unit cells, where different values of $\eta$ and $\delta_x$ are chosen. One of the non-Hermitian parameters $\delta_y$ is fixed to be $0.05$.}
  \label{fig7}
\end{figure}

In this section, we study the competition between the PMF and NHSE when the skin modes are localized at the corner. To this end, we consider a finite lattice with $40\times40$ unit cells. The parameter $\delta_y$ is fixed to be 0.05 while $\delta_x$ and $\eta$ are tunable. To demonstrate the suppression of skin modes, we investigate the evolution of an eigenstate belonging to the first-order Landau level in the Hermitian limit. As illustrated in Fig.~\ref{fig7}, the PMF indicator $\eta$ and the other non-Hermitian parameter $\delta_x$ are varied accordingly to highlight the competition.

When $\delta_x$ and $\eta$ are both zero, the eigenstate localizes at the top edge as expected from nonreciprocal couplings along $y$ direction. By increasing $\delta_x$, the skin mode distributed along the top edge is then pushed to the top right corner, thus inducing a corner skin mode. Starting from this corner skin mode (i.e., the bottom left panel in Fig.~\ref{fig7}), as can be seen, by varying $\eta$ and $\delta_x$, the skin mode can be driven along both $x$ and $y$ directions, showing a large degree of freedom in the skin mode manipulation in this finite lattice. In particular, with increasing $\eta$ and decreasing $\delta_x$, the corner skin mode is gradually transferred to a bulk Landau mode (see the top right panel in Fig.~\ref{fig7}).

\begin{figure*}
  \centering
  \includegraphics[width=\linewidth]{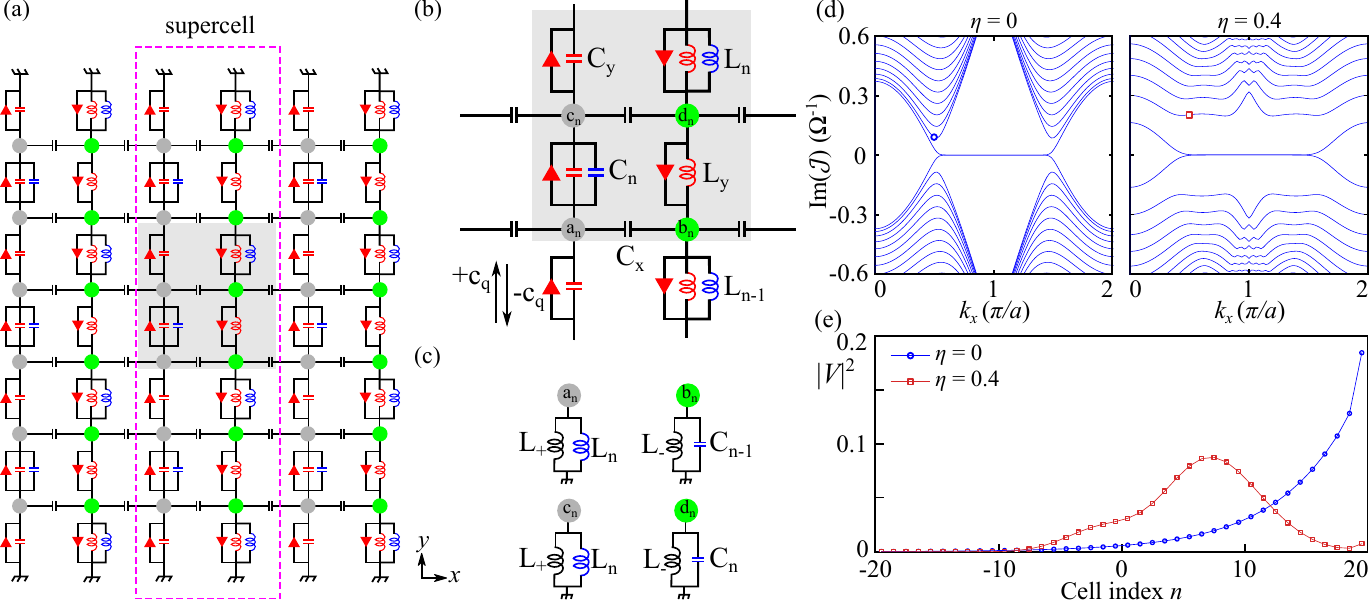}
  \caption{Panels (a) to (c) depict the electric circuit design for realizing our scheme. For visualization, only three unit cells along the $y$ direction are shown in panel (a). The inductors and capacitors are represented by $L$ and $C$, respectively. $\pm c_q$ are the directional dependent capacitance of the INICs represented by the red triangles. (d) Admittance spectra of the circuit with $N_y=40$, while considering different PMF strengths $\eta=0$ and $0.4$, respectively. $C_x=1~\mu$F and $c_q=50$~nF fix the non-Hermitian parameters as $(\delta_x,\delta_y)=(0,0.05)$. (e) Voltage variation of an eigenstate near zero admittance at the $k_x=\pi/2$ valley, as labelled in panel (d) with filled markers at different $\eta$.}
  \label{fig8}
\end{figure*}

\section*{Appendix C: Proposal for electric circuit realization}
Topoelectrical circuit is an excellent platform for realizing NHSE-based phenomena. Based on previous developments in this field \cite{imhof2018topolectrical, helbig2020generalized, liu2021non, zou2021observation}, we hereby demonstrate a circuit realization of the lattice illustrated in Fig.~\ref{fig1}. The reciprocal part of the circuit consists of inductors and capacitors; whereas the nonreciprocity is achieved via negative impedance converters with current inversion (INICs) \cite{imhof2018topolectrical}. A schematic diagram of such a circuit in a strip geometry is shown in Fig.~\ref{fig8}(a), having $x$ axis as the periodic direction and consisting of three unit cells along the $y$ axis. The nodes are shown in gray and green circles, which play the role of the sites. A zoomed view of the $n$th cell within the supercell is shown in Fig.~\ref{fig8}(b), which consists of four nodes $(a_n,b_n,c_n,d_n)$. The positive couplings along the $x$ axis are realized by the capacitors (in black) $C_x$ and those along the $y$ axis are realized by the capacitors (in red) $C_y$. The negative couplings are realized by the proper choice of the inductors $L_y$. The INICs (represented by the red triangles) are connected in parallel with $C_y$ and $L_y$. However, the directions of the INICs are reversed while connecting with $L_y$. Additional inductors and capacitors $(L_n,C_n)$ are connected properly in order to realize the PMF as labelled in blue in Fig.~\ref{fig8}(b). All the nodes in the circuit are grounded as shown in Fig.~\ref{fig8}(c).

The role of the Hamiltonian $\mathcal{H}$ is played by the Laplacian $\mathcal{J}$ of the circuit with $\mathcal{J}\propto -i\mathcal{H}$ and $I=\mathcal{J} V$, where $I$ and $V$ are the vectors representing the currents and voltages at each node. Following Ref.~\cite{lee2018topolectrical}, the currents at each node of the $n$th cell can be expressed as:
\begin{widetext}
\begin{align}
I_{a_n} =&\left[\frac{1}{i\omega}\left( L_+^{-1}+ L_n^{-1}\right)+i\omega(2C_x+2C_y+C_n)\right]V_{a_n}-i\omega(C_y+c_q)V_{c_{n-1}}-i\omega(C_y+C_n-c_q)V_{c_{n}}\notag\\
&-i\omega C_x(1+e^{-ik_xa})V_{b_{n}},\\
I_{b_n} =&\left[ (i\omega L_-)^{-1}+i\omega(C_{n-1}+2C_x)+2(i\omega L_y)^{-1}+(i\omega L_{n-1})^{-1}\right]V_{b_n}-\left[\frac{1}{i\omega}\left( L_y^{-1}+ L_{n-1}^{-1}\right)-i\omega c_q\right] V_{d_{n-1}}\notag\\
&-\left[\left(i\omega L_y\right)^{-1}+i\omega c_q\right] V_{d_{n}}-i\omega C_x(1+e^{ik_xa})V_{a_{n}},\\
I_{c_n} =&\left[\frac{1}{i\omega}\left( L_+^{-1}+ L_n^{-1}\right)+i\omega(2C_x+2C_y+C_n)\right]V_{c_n}-i\omega(C_y+C_n+c_q)V_{a_{n}}-i\omega(C_y-c_q)V_{a_{n+1}}\notag\\
&-i\omega C_x(1+e^{-ik_xa})V_{d_{n}},\\
I_{d_n} =&\left[ (i\omega L_-)^{-1}+i\omega(C_{n}+2C_x)+2(i\omega L_y)^{-1}+(i\omega L_{n})^{-1}\right]V_{d_n}-\left[\left(i\omega L_y\right)^{-1}-i\omega c_q\right] V_{b_{n}}\notag\\
&-\left[\frac{1}{i\omega}\left( L_y^{-1}+ L_{n}^{-1}\right)+i\omega c_q\right]V_{b_{n+1}}-i\omega C_x(1+e^{ik_xa})V_{c_{n}}.
\end{align}
\end{widetext}
Here $\omega=2\pi f_0$, where $f_0$ is the resonant frequency of the circuit. In a finite circuit, to realize $y$-OBC, the nodes at the edges are grounded properly, whereas $x$-PBC can be realized by connecting the two edges along the $x$ direction through conducting wires. In order to realize the PMF, we choose: 
\begin{align}
C_n=&2C_x\cos{\left[\frac12\left(\frac\pi2-\frac{2\pi n}{N_y}\eta \right)\right]},\notag\\
L_n=&1/(\omega^2C_n).
\end{align}

We aim for resonant frequency $f_0=100$~kHz  and choose the values of the circuit components accordingly: $C_x=1~\mu$F, $C_y=C_x/2$, $L_y=1/(\omega^2C_y)\approx5~\mu$H, $L_+=1/(6\omega^2C_y)\approx0.8~\mu$H, $L_-=L_y/2$, and $c_q=50$~nF. At the resonant frequency, all the diagonal terms of the circuit Laplacian $\mathcal{J}$ become zero. In Fig.~\ref{fig8}(d), the band structures without and with the PMF are presented, where the appearance of the Landau levels for $\eta\neq0$ is clearly shown. Such a band structure can be measured in practice from the admittance response. Figure~\ref{fig8}(e) shows the skin effect suppression due to the PMF, which can be obtained in practice by measuring the voltage across all the nodes of the circuit.

\end{document}